\documentclass[reprint,amsmath,amssymb,aps,twocolumn,longbibliography,showpacs]{revtex4}

\usepackage{graphicx}
\usepackage{dcolumn}
\usepackage{bm}
\usepackage{textcomp}

\begin{document}

\preprint{APS/123-QED}

\title{MHD Equilibrium Equation in Symmetric Systems}
\thanks{Originally published by {\it PUBLICA\c{C}\~OES IFUSP/P-817}}

\author{Mutsuko Y. Kucinski}
\author{Iber\^e L. Caldas}
 \altaffiliation{Partially Supported by CNPq}
\affiliation{Instituto de F\'isica, Universidade de S\~ao Paulo, P.O. Box 66318, S\~ao Paulo, Brasil}

\begin{abstract}
In MHD symmetric systems the equilibrium physical quantities are dependent on two variables only. In this cases it is possible to find a magnetic surface function that has the same symmetry. Under the assumption that the metric determinant is also independent of a third, ignorable coordinate, a general MHD equilibrium equation in curvilinear coordinates is deduced. This equation is specially useful when non-orthogonal generalized coordinates are used.
\end{abstract}


\maketitle

\section{Introduction}
Ideal magnetohydrodynamics (MHD) is the most basic single-fluid model for determining the macroscopic equilibrium and stability properties of a plasma \cite{wesson1979,freidberg1982}. There is enough evidence that this model describes how magnetic, inertial and pressure forces interact within a perfectly conducting plasma \cite{bateman1978,goedbloed1979}.

The ideal MHD is used in magnetic fusion to describe static equilibria and to infer convenient magnetic geometries for confinement. In symmetric plasma systems the field lines lie on a set of closed nested toroidal magnetic surfaces. These surfaces can be determined by solving the MHD equilibrium equation \cite{greene1965,grad1985}.

Equilibrium equations for different symmetric plasma systems have appeared in the literature \cite{freidberg1982}. In particular, the well known Grad-Shafranov \cite{hain1957,freidberg1982} equation is written in terms of orthogonal coordinates and describes axisymmetric toroidal equilibrium.

In this article a general MHD symmetric system is described by a magnetic surface with the same symmetry that the considered equilibrium. This function is obtained from a general equilibrium equation in curvilinear coordinates. With this equation, once the curvilinear coordinates has been chosen, the equilibrium equation in any geometry can be derived. This is especially useful when non-orthogonal generalized coordinates are used.

The general curvilinear coordinates used in this article are defined in Section I. A symmetric transversal magnetic flux $\Psi$ and a function $I$ that determines the transversal electric current are introduced in Sections II and III. In Section IV, a general equilibrium equation relating the equilibrium pressure to the surface functions $\Psi$ and $I$ is derived. Finally systems well-known in the literature are considered as examples and the equilibrium equations in terms of conventional toroidal coordinates (Appendix A), helical coordinates (Appendix B) and natural coordinates (Appendix C) are derived from the general equation obtained in this article.

\section{Curvilinear Coordinates}
In symmetric plasma confinement systems, all the equilibrium functions having a physical meaning are dependent on two variables only. Curvilinear coordinates are named $u_1,u_2$ and $u_3$. The surfaces $u_i=c_i$, where $c_i$ is a constant, are coordinate surfaces. A coordinate curve $u_k$ is a curve along which $u_i, u_j$ ($i\neq j\neq k$) are constants.

The coordinates $u_1$ and $u_2$ are chosen in order to have the magnetic axis of the system coincident with a coordinate curve $u_3$ and $u_2$ is a transversal coordinate, $u_3$ will be an ignorable coordinate; longitudinal directions are given by coordinate curves $u_1$. In plasma confinement problems we have, usually, periodicity in $u_2$ and $u_3$. The following periodicity
\begin{equation}\label{eq01}
 u_3=L(u_1,u_2)
\end{equation}
is assumed.

An attempt is made use of the notations most familiar in the literature.

The covariant basis vectors are given by
\begin{equation}\label{eq02}
 \bm e_i=\frac{\partial \bm r}{\partial u_i}
\end{equation}
where $\bm e_i$ is tangent to the $u_i$ curve and the contravariant basis vectors are defined by
\begin{equation}\label{eq03}
 \bm e^i=\nabla u_i
\end{equation}
where $\bm e^i$ is normal to the $u_i$ surface. $u_1$,$u_2$ and $u_3$ are taken in order to satisfy
\begin{equation}\label{eq04}
 \bm e_i=\sqrt{g}\bm e^j\times e^k
\end{equation}
for any cyclic permutation $(i,j,k)$, $g$ is the determinant of the covariant metric tensor $g_{i,j}$.

\section{Transversal Magnetic Flux}
Define $L\Psi(u_1,u_2)$ as the magnetic flux through a coordinate surface $u_2$ which extends from the magnetic axis to a coordinate curve $u_3$ and limited by $0\leq u_3\leq L$. On the magnetic axis $u_1=a$ and
\begin{equation}\label{eq05}
 \bm B=B^3\bm e_3
\end{equation}
Then:
\begin{equation}\label{eq06}
 L\Psi=\int_a^{u_1}du'_1\int_0^L\sqrt{g}B^2du_3
\end{equation}
from which it follows
\begin{equation}\label{eq07}
 \frac{\partial \Psi}{\partial u_1}=\frac{1}{L}\int_0^L\sqrt{g}B^2du_3
\end{equation}
Taking account of the equation
\begin{equation}\label{eq08}
 \nabla\cdot\bm B=0
\end{equation}
and assuming $B^1=0$ on the axis we can derive from expression (\ref{eq06}):
\begin{equation}\label{eq09}
 \frac{\partial\Psi}{\partial u_2}=-\frac{1}{L}\int_0^L \sqrt{g}B^1du_3.
\end{equation}
If $\sqrt{g}B^1$ and $\sqrt{g}B^2$ are independent of $u_3$, we find an expression for $B$ in terms of $\Psi$:
\begin{equation}\label{eq10}
 \bm B=\frac{\bm e_3}{g_{33}}\times\nabla\Psi+B_3\frac{\bm e_3}{g_{33}}
\end{equation}
$\Psi=\mbox{constant}$ represents a magnetic surface because
\begin{equation}\label{eq11}
 \nabla\Psi\cdot\bm B=0
\end{equation}
as can be seen using (\ref{eq10}).

The magnetic flux can also be expressed in terms of the vector potential $\bm A$ using the Stokes theorem:
\begin{equation}\label{eq12}
 \Psi=-\frac{1}{L}\int_0^L A_3du_3,
\end{equation}
$A_3$ is assumed to be zero on the axis. In symmetric systems:
\begin{equation}\label{eq13}
 \Psi(u_1,u_2)=-A_3(u_1,u_2).
\end{equation}
The only restriction to the gauge of $\bm A$ is in order to keep the same symmetry as the physical quantities.

\section{Current Density}
The current density satisfies the equations
\begin{equation}\label{eq14}
 \nabla\times\bm B=\mu_0\bm J \mbox{ and } \nabla\cdot\bm J=0.
\end{equation}
On the magnetic axis $J^1=0$. These equations are similar to the equations for $B$:
\begin{equation}\label{eq15}
 \nabla\times\bm A=\bm B \mbox{ and } \nabla\cdot\bm B=0
\end{equation}
and also $B^1=0$ on the axis.

Similar considerations must yield similar results. Let us define a function:
\begin{equation}\label{eq16}
 \mu_0 I=-\frac{1}{L}\int_0^LB_3du_3,
\end{equation}
which is an expression similar to (\ref{eq12}). In a symmetric case it would be:
\begin{equation}\label{eq17}
 \mu_0 I=-B_3(u_1,u_2).
\end{equation}
In this case on the axis $B_3$ is not zero.

The transversal current is given by:
\begin{equation}\label{eq18}
 L(I-I_{axis})=\int_\sigma \sqrt{g}J^2 du_1du_3.
\end{equation}
This expression can be compared to (\ref{eq06}). Thus considering the symmetry argument, the current density is expressed in terms of $I$ as:
\begin{equation}\label{eq19}
 \bm J=\frac{\bm e_3}{g_{33}}\times\nabla I+J_3\frac{\bm e_3}{g_{33}},
\end{equation}
an expression similar to (\ref{eq10}).

\section{Pressure Equilibrium Equation}
The MHD equilibrium theory, with scalar pressure $P$, considers the equations:
\begin{equation}\label{eq20}
 \nabla P=\bm{J}\times\bm B
\end{equation}
and
\begin{equation}\label{eq21}
 \nabla\times\bm B=\mu_0\bm J,
\end{equation}
$I, P$ and $\Psi$ satisfies the relations:
\begin{eqnarray}
 \bm B\cdot\nabla\Psi=0 & , & \bm B\cdot\nabla P=0\\
 \bm J\cdot\nabla P=0 & , & \bm J\cdot\nabla I=0,
\end{eqnarray}
what means that $I, P$ and $\Psi$ are surface quantities.

Using (\ref{eq20}) together with (\ref{eq17}) and (\ref{eq19}) a relation between these surface quantities is found:
\begin{equation}\label{eq24}
 \nabla P=-\frac{J_3}{g_{33}}\nabla\Psi+\frac{B_3}{g_{33}}\nabla I,
\end{equation}
$J_3$ can be taken from (\ref{eq21}) and $B_3=-\mu_0I$. Substituting them in (\ref{eq24}) and using (\ref{eq10}) we find the final expression:
\begin{eqnarray}
 &&(\Delta^*\Psi)\nabla \Psi=-\mu_0g_{33}\nabla P-\mu_0^2I\nabla I\nonumber\\
&&+\mu_0I\frac{g_{33}}{\sqrt{g}}\left[\frac{\partial}{\partial u_1}\left(\frac{g_{23}}{g_{33}} \right)-\frac{\partial}{\partial u_2}\left(\frac{g_{13}}{g_{33}}\right) \right]\nabla\Psi \label{eq25},
\end{eqnarray}
where
\begin{eqnarray}
 &&\Delta^*\Psi\equiv \frac{g_{33}}{\sqrt{g}}\left[\frac{\partial}{\partial u_1}\frac{\sqrt{g}}{g_{33}}\left(g^{11}\frac{\partial\Psi}{\partial u_1}+g^{12}\frac{\partial\Psi}{\partial u_2} \right)\right.\nonumber\\
&&\left.+\frac{\partial}{\partial u_2}\frac{\sqrt{g}}{g_{33}}\left(g^{12}\frac{\partial\Psi}{\partial u_1}+g^{22}\frac{\partial\Psi}{\partial u_2}\right)\right]\label{eq26}
\end{eqnarray}
$I$ and $P$ are functions of $\Psi$ only. Therefore, whenever $\nabla\Psi\neq 0$, the expression can be simplified to a scalar equation
\begin{eqnarray}
 &&\Delta^*\Psi=-\mu_0g_{33}P'-\mu_0^2II'\nonumber\\
&&+\mu_0I\frac{g_{33}}{\sqrt{g}}\left[\frac{\partial}{\partial u_1}\left(\frac{g_{23}}{g_{33}} \right)-\frac{\partial}{\partial u_2}\left(\frac{g_{13}}{g_{33}}\right)\right]\label{eq27}
\end{eqnarray}
which corresponds to the Grad-Shafranov equation \cite{freidberg1982}. Here the prime indicates differentiation with respect to $\Psi$. The quantity $\mu_0g_{33}P+\mu_0^2I^2/2$ must be continuous through a surface where $\nabla\Psi=0$.

The components of the magnetic field are expressed in terms of $\Psi$ as:
\begin{eqnarray}
 \sqrt{g}B^2&=&\frac{\partial\Psi}{\partial u_1}\nonumber\\
&\mbox{and}& \\
 \sqrt{g}B^1&=&-\frac{\partial\Psi}{\partial u_2}\nonumber
\end{eqnarray}

The equation (\ref{eq26}) can also be written as:
\begin{eqnarray}
 \Delta^*\Psi&\equiv&\nabla^2\Psi-\nabla\Psi\cdot\frac{\nabla g_{33}}{g_{33}}\\
 &\mbox{or}& \nonumber\\
 \Delta^*\Psi&\equiv&g_{33}\nabla\cdot\left(\frac{\nabla\Psi}{g_{33}}\right)
\end{eqnarray}
The general equilibrium equation can be specially useful when non-orthogonal generalized coordinates are used.

In Appendix A toroidal pinches with axial symmetry are considered, and the equilibrium equation in conventional toroidal coordinates \cite{shafranov1960} is derived from equation (\ref{eq14}). If the system presents a straight helical symmetry, the coordinates $u_1=r$, $u_2=\theta-\alpha z$ and $u_3=z$ can be introduced, where $\alpha$ is the pitch of the helix. Using (\ref{eq25}) the equilibrium equation as found on the literature \cite{freidberg1982} is deduced straightforwardly. In these systems it is very likely to appear discontinuity surfaces where $\nabla\Psi=0$ (See Appendix B). The well-known equilibrium equation using flux coordinates \cite{freidberg1982} is obtained also very easily in Appendix C.

\section{Conclusions}
In this article a general MHD symmetric system has been considered, and a magnetic surface function, with the same symmetry, has been introduced to describe it. An equilibrium equation satisfied by this function is deduced in curvilinear coordinates. This equation is a generalization of several particular MHD equilibrium equations valid for the equilibria considered in the literature. Therefore, given the curvilinear coordinates, the equilibrium equation for a magnetic surface function, in any geometry, can be derived. This is especially useful if nonorthogonal coordinates are used. As examples, in the appendices, equilibrium equations well known in the literature are obtained from the mentioned general equation presented in this article. The procedure followed in this paper resembles the one used to deduce an equilibrium equation for incompresible irrotational steady fluid flow \cite{sparenberg1989}.

\appendix
\section{Toroidal pinch with axial symmetry}
\begin{figure}[h]
 \centering
 \includegraphics[width=0.4\textwidth]{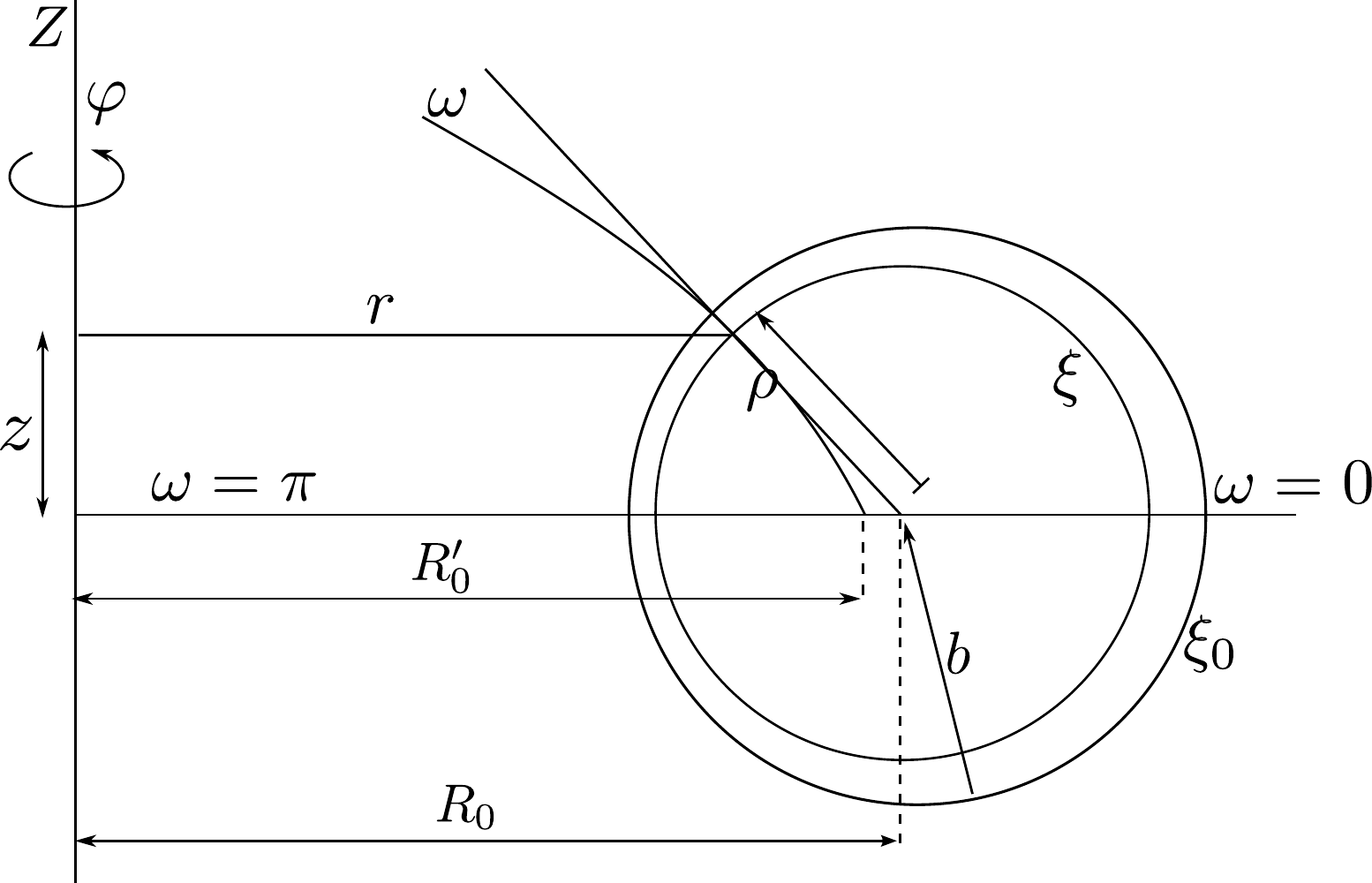}
 \caption{\label{figure}Toroidal coordinate system.}
\end{figure}

The equilibrium equation using a conventional toroidal coordinate system \cite{shafranov1960} (see Fig.\ref{figure}).
\begin{equation}\nonumber
 u_1=\xi\mbox{ , }u_2=\omega\mbox{ , }u_3=\varphi
\end{equation}
is obtained in this appendix. The coordinates are defined by:
\begin{equation}\nonumber
 r=\frac{R_0'\sinh\xi}{\cosh\xi-\cos\omega}\mbox{ , }z=\frac{R_0'\sin\omega}{\cosh\xi-\cos\omega}
\end{equation}
were $r,z$ and $\varphi$ are the polar cylindrical coordinates and $R_0$ is the major radius. If $\xi=\xi_0$ defines the toroidal surface, then $\cosh\xi_0=R_0/b$, $R_0'=R_0\sqrt{1-b^2/R_0^2}$.

The contravariant basis is

\begin{eqnarray}
 \bm e^1=\nabla\xi=\frac{\bm e_\xi}{h_\xi}\nonumber\\
 \bm e^2=\nabla\omega=\frac{\bm e_\omega}{h_\omega}\nonumber\\
 \bm e^3=\nabla\varphi=\frac{\bm e_\varphi}{h_\varphi}\nonumber
\end{eqnarray}
with
\begin{equation}\nonumber
 h_\xi=h_\omega=\frac{R_0'}{\cosh\xi-\cos\omega}
\end{equation}
and,
\begin{equation}
 h_\varphi=h_\xi\sinh\xi\nonumber
\end{equation}
$\bm e_\xi,\bm e_\omega$ and $\bm e_\varphi$ are unit vectors. The metric is given by:
\begin{eqnarray}
\sqrt{g}=(\bm e^1\cdot\bm e^2\times \bm e^3)^{-1}=h_\xi h_\omega h_\varphi\nonumber\\
g^{11}=\frac{1}{h_\xi^2}=g^{22}\mbox{ ; }g_{33}=h_\varphi^2.\nonumber
\end{eqnarray}
The equilibrium equation (\ref{eq27}) becomes:
\begin{equation}
  \frac{\sinh\xi}{h_\xi}\left( \frac{\partial}{\partial\xi}\frac{1}{h_\varphi}\frac{\partial\Psi}{\partial\xi}+\frac{\partial}{\partial\omega}\frac{1}{h_\varphi}\frac{\partial\Psi}{\partial\omega} \right)=-\mu_0h_\varphi^2P'-\mu_0^2 II'
\end{equation}
If a function $F$ is introduced as:
\begin{equation}
 \Psi=[2(\cosh\xi-\cos\omega)]^{-1/2}F
\end{equation}
we obtain the well known Grad-Shafranov equation \cite{freidberg1982}:
\begin{eqnarray}
 &&\frac{\partial^2F}{\partial\xi^2}+\frac{\partial^2F}{\partial\omega^2}-\coth\xi\frac{\partial F}{\partial\xi}+\frac{1}{4}F=-\frac{2R_0'^2}{[2(\cosh\xi-\cos\omega)]^{3/2}}\nonumber\\
&&\times\left\{\frac{4R_0'^2\sinh^2\xi}{[2(\cosh\xi-\cos\omega)]^2}\mu_0P'+\mu_0^2II'\right\}
\end{eqnarray}
The magnetic field is derived using (\ref{eq27}):
$$
 h_{\xi}h_{\varphi}B_\omega=\frac{\partial \Psi}{\partial \xi}
$$
and
\begin{equation}
 h_{\omega}h_{\varphi}B_\xi=-\frac{\partial \Psi}{\partial \omega}
\end{equation}

\section{Helical system with a straight magnetic axis}
The coordinates are:
\begin{equation}\nonumber
 u_1=r\mbox{ , }u_2=\theta-\alpha z\equiv u\mbox{ , }u_3=z
\end{equation}
where $r,\theta$ and $z$ are the polar coordinates. Their contravariant basis is:
\begin{equation}\nonumber
 \bm e^1=\bm e_r{ , }\bm e^2=\frac{\bm e_\theta}{r}-\alpha\bm e_z\mbox{ , }\bm e^3=\bm e_z
\end{equation}
where $\bm e_r,\bm e_\theta$ and $\bm e_z$ are the unit vectors in cylindrical coordinates, the covariant basis is:
\begin{equation}\nonumber
 \bm e_1=\bm e_r\mbox{ , }\bm e_2=r\bm e_\theta\mbox{ , }\bm e_3=\bm e_z+\alpha r\bm e_\theta
\end{equation}
The metric is given by:
$$
g^{ij}=\left[
\begin{array}{ccc}
 1&0&0\\
 0&\frac{1}{r^2}+\alpha^2 &-\alpha\\
 0&-\alpha&1
\end{array}
\right]
\mbox{ , }
g_{i,j}=
\left[
\begin{array}{ccc}
 1&0&0\\
 0&r^2&\alpha r^2\\
 0&\alpha r^2&1+\alpha^2r^2\\
\end{array}
\right]
$$
In this case the equilibrium equation (\ref{eq27}) becomes \cite{freidberg1982}
\begin{eqnarray}
 &&\frac{1+\alpha^2r^2}{r}\left[\frac{\partial}{\partial r}\left(\frac{r}{1+\alpha^2 r^2}\frac{\partial\Psi}{\partial r}\right)+\frac{1}{r}\frac{\partial^2\Psi}{\partial u^2}\right]=\nonumber\\
 &&-(1+\alpha^2r^2)\mu_0P'-\mu_0^2II'+\mu_0I\frac{1+\alpha^2r^2}{r}\frac{\partial}{\partial r}\frac{\partial r^2}{1+\alpha^2r^2}\nonumber\\
 &&
\end{eqnarray}
and the magnetic field components are derived using the following equations:
\begin{equation}
 \frac{\partial \Psi}{\partial r}=B_\theta -\alpha rB_z \mbox{ , } \frac{\partial \Psi}{\partial u}=-rB_r
\end{equation}

\section{Natural coordinates}
In natural systems \cite{hamada1962} the coordinate $u_1$ is a magnetic surface label, it is analogous to the minor radius of the torus, $u_2$ and $u_3$ are poloidal and toroidal cyclic coordinates, ranging from $0$ to $2\pi$.

The physical variables are:

\vspace{0.5cm}
\begin{tabular}{l l}
 Magnetic pressure & $P(u_1)$\\
 Poloidal flux & $2\pi\chi(u_1)=\int_0^{u_1}du_1'\int_0^{2\pi}\sqrt{g}B^2du_3$\\
 Poloidal current & $2\pi(I-I_{axis})=\int_0^{u_1}du_1'$\\
                  & $\times\int_0^{2\pi}\sqrt{g}J^2du_3$\\
 Toroidal flux & $2\pi\Phi=\int_0^{u_1}du_1'\int_0^{2\pi}\sqrt{g}B^3du_2$\\
 Toroidal current & $2\pi J=\int_0^{u_1}du_1'\int_0^{2\pi}\sqrt{g}J^3du_2$\\
 Volume inside a & $(2\pi)^2V(u_1)=\int_0^{u_1}du_1'\int_0^{2\pi}du_2$\\
 toroidal magnetic& $\times\int_0^{2\pi}\sqrt{g}du_3$\\
 surface &\\
\end{tabular}
\vspace{0.5cm}

The equilibrium equation (\ref{eq24}) becomes:
\begin{equation}\label{eqc1}
 P'=-\frac{J_3}{g_{33}}\chi'-\frac{\mu_0I}{g_{33}}I'
\end{equation}
where the prime indicates derivation with respect to $u_1$. Multiplying equation (\ref{eqc1}) by $\sqrt{g}du_2du_3$ and integrating in a magnetic surface we get
\begin{equation}\label{eqc2}
 P'V'=-\chi'J'-\Phi'I'
\end{equation}
The magnetic field and the current density are given by:
\begin{equation}\label{eqc3}
 \bm B=\frac{\bm e_3}{g_{33}}\times\nabla\Psi-\frac{\mu_0I}{g_{33}}\bm e_3
\end{equation}
and
\begin{equation}\label{eqc4}
 \bm J=\frac{\bm e_3}{g_{33}}\times\nabla I+\frac{J_3}{g_{33}}\bm e_3
\end{equation}
In terms of the contravariant components (\ref{eqc4}) becomes:
\begin{equation}\label{eqc5}
 \bm J=\frac{I'}{\sqrt{g}}\bm e_2+J^3\bm e_3
\end{equation}
Introducing the average value of $J^3\sqrt{g}$ in a magnetic surface:
$$
\langle J^3\sqrt{g}\rangle=\frac{1}{2\pi}\int_0^{2\pi}J^3\sqrt{g}du_2
$$
and writing
$$
J^3\sqrt{g}=\langle J^3\sqrt{g}\rangle+\overline{J^3\sqrt{g}}
$$
expression (\ref{eqc5}) becomes:
$$
\bm J=\frac{I'}{\sqrt{g}}\bm e_2+\frac{J'}{\sqrt{g}}\bm e_3+\frac{\overline{J^3\sqrt{g}}}{\sqrt{g}}\bm e_3
$$

Defining a function $\nu$ by:
$$
\frac{\partial \nu}{\partial u_2}=\overline{J^3\sqrt{g}}
$$
the current density can be written:
$$
\bm J=\nabla\times(-I\bm e^3+J\bm e^2-\nu\bm e^1)
$$
and $\bm B$ comes naturally as:
$$
\frac{\bm B}{\mu_0}=-I\bm e^3+J\bm e^2-\nu\bm e^1+\nabla\Phi
$$
where $\Phi$ is the scalar potential in the absence of the plasma. This equation equated to (\ref{eqc3}) represents the equations commonly used to determine the flux coordinates \cite{hamada1962,hirshman1982}.

The only assumption taken is of the symmetry of the system.

\newpage

\bibliography{references}

\end{document}